\documentstyle[11pt,IAUS215,twoside,epsf]{article}

\def\edcomment#1{\iffalse\marginpar{\raggedright\sl#1\/}\else\relax\fi}
\reversemarginpar

\begin{document} 
\vspace*{1cm} 
\title{White Dwarf Rotation: Observations and Theory} 
\author{Steven D. Kawaler} 
\affil{Department of Physics and Astronomy,
Iowa State University, Ames, IA  50011 USA}

\begin{abstract} 
White dwarfs rotate.  The angular momentum in single white
dwarfs must originate early in the life of the star, but also must be modified
(and perhaps severely modified) during the many stages of evolution between
birth as a main--sequence star and final appearance as a white dwarf.
Observational constraints on the rotation of single white dwarf stars come
from traditional spectroscopy and from asteroseismology, with the latter
providing hints of angular velocity with depth.  Results of these
observational determinations, that white dwarfs rotate with periods ranging
from hours to days (or longer), tells us that the processes by which angular
momentum is deposited and/or drained from the cores of AGB stars are complex. 
Still, one can place strong limits on these processes by considering
relatively simple limiting cases for angular momentum evolution in prior
stages, and on subsequent angular momentum evolution in the white dwarfs. 
These limiting-case constraints will be reviewed in the context of the
available observations.

\end{abstract}

\section{Introduction}

Despite a flurry of interest, mainly from theorists, in the 1960s, applicable
methods to observe rotation in single white dwarfs largely eluded astronomers
until the 1970s.  As relatively simple structures whose mechanics are
dominated by the degenerate equation of state of the electrons, the influence
of rotation on their structure can, in principle, be computed analytically
using the theory of rotating polytropes.  Rotation could provide another
``force'' that could support the star, increasing the maximum mass for an
electron degenerate configuration.   Thus the hope was, in part, that by
finding white dwarfs rotating at sufficiently high velocities, physics at the
interface between classical and modern could be tested. One of the earliest
efforts to explore this connection was by Ian Roxburgh (Roxburgh 1965), who
showed that the maximum mass of a white dwarf could be increased by about 5\%
if rotating at near critical velocity of over 2000 km/s. As limits on rotation
velocities of single white dwarfs began to be measured in the 1970s, it became
apparent that they were rotating at much smaller velocities than that.  Still,
measuring the rate of rotation of white dwarfs can allow us to probe the
evolution of stars in their late stages -- the rotation that we do see is a
remnant of the initial angular momentum, processed by several stages of
stellar evolution.

\section{Expectations from Scaling Arguments - Fast versus Slow}

One curiosity instantly arises in the study of white dwarfs rotation -- that
is, what should be considered a ``fast'' rotation velocity, and what is a
``slow'' rotation rate?  As we will see below, measured white dwarf rotation
rates seem to cluster around once per day.  The literature contains examples
of papers that talk about such rates as being fast, and others that refer to
them as slow.  This ambiguity stems from the two extreme cases of angular
momentum redistribution that bracket what actually happens in stars.  Assuming
that local angular momentum (hereafter ``$j$'') is conserved through evolution
results in an estimate of rotation at a rate of several hundred km/s. 
Alternatively, assuming that stars always rotate as solid bodies leads to an
estimate of very long rotation periods (following loss of the extended AGB
envelope).  The view that even small magnetic fields can enforce solid--body
rotation in all phases of evolution (i.e. Spruit 1998) leads directly to this
sort of result.  

Let's see more quantitatively what these two extreme cases imply.  Assume that
a 0.6$M_{\odot}$ white dwarf star begins its life on the main sequence with a
total mass of 3$M_{\odot}$ and total angular momentum consistent with the
Kraft relation (i.e. Kawaler 1987 and references therein).  Thus the initial
rotation period is approximately 10 hours.

If $j$ is retained throughout the
evolution, internal differential rotation must develop, as parts of
the star expand or shrink through evolution.  By ignoring the rotational
instabilities that should result from such angular velocity gradients, we can
obtain an upper limit to the rotation rate of white dwarfs.  The central
0.6$M_{\odot}$ of a 3$M_{\odot}$ main--sequence star collapse by a factor of
about 20 by the time the star eventually reaches white dwarf dimensions.  Thus
it will spin up by approximately a factor of 400, have a rotation period of
about 100 seconds, and a surface equatorial rotation velocity of 680 km/s. 
Under this assumption, then, anything slower than many hundreds of km/s for a
rotation velocity would be considered slow rotation.  

If the white dwarf evolves from a lower mass progenitor, that
the rotation velocity could be much lower even in this case.  Stars of
main--sequence spectral class of late F or later can lose up to 99\% of their
angular momentum before and on the main sequence.  If they conserve $j$, their
rotation rates as white dwarfs could be as small as about 10 km/s.  But, given
the demographics of white dwarfs (a mean mass of about 0.6$M_{\odot}$ and
Population I kinematics) suggest that most that we see are derived from stars
above the break in the Kraft curve, and therefore should have rotation
velocities of hundreds of km/s if they are described by the assumption of
locally conserved angular momentum.

Next, consider a star that forever rotates as a solid body.  When it reaches
the AGB, its radius will have swelled to 90 times its main--sequence radius.  
It will then be rotating at a rate that is roughly 7500 times slower than its
main--sequence rate,  with nearly all of the angular momentum  seated in the
outer layers at large radius.  The rotation period is 8.5 years.   Upon loss
of the envelope and exposure of the central 0.6$M_{\odot}$ white dwarf, the
rotation period may shorten by a factor of 4 as the remnant contracts, leaving
a white dwarf with a rotation period of about 2 years and a vanishingly small
equatorial rotation velocity of about 1 meter per second.  This assumption
gives an {\em upper} limit to the rotation velocity, as stars must lose
angular momentum as they lose mass on the AGB.  Thus, under the assumption of
perpetual solid--body rotation, any measurable rotation in a white dwarf
should be considered fast rotation!  

The fact that some white dwarfs are seen to rotate argues that some other
mechanism is needed to be invoked to impart angular momentum to the remnant
(e.g. Spruit 1998) in this extreme.  Spruit (1998) points out that even small
magnetic fields should enforce solid--body rotation in all phases of evolution
-- and that they should show therefore no measurable rotation. The fact that
they are indeed seen to rotate suggests a different origin entirely outside of
the primordial angular momentum.  He argues that some white dwarfs may in fact
rotate because of asymmetric mass loss on the AGB phase, and during the final
superwind phase.  Though this is possible, we outline below another origin for
the observed angular velocities in white dwarfs.

\section{Observations of White Dwarf Rotation}

Traditional measurements of rotation rates for stars -- photometric variations
from spots and rotational broadening of otherwise narrow absorption lines --
fail for white dwarfs.  White dwarfs are spotless stars -- photometric time
series do not show evidence of rotational modulation by spots.  With very high
gravities producing lines that are 10 nm wide or more, natural broadening
mechanisms swamp rotational broadening -- even for velocities close to
breakup.  However, discovery of sharp cores in the spectral lines of some DA
white dwarfs, and of nonradial pulsations that are influenced in a predictable
way by rotation, have produced a handful of useful (single) white dwarf
rotation velocities.  Magnetic white dwarfs, with magnetic fields in excess of
100 kG, show evidence of time--variability (and in some cases periodicity) of
the magnetic features, providing a measurement of their rotation velocities. 
Given that such magnetic white dwarfs are rare (approximately 1\% of white
dwarfs) and that the magnetic fields can influence the rotational evolution,
we do not discuss them further in this review -- but see the excellent review
by Wickramasinghe \& Ferrario (2000) for more on these fascinating objects.

\subsection{Spectroscopic Measurements of $v \sin i$ in White Dwarfs}

It was not until the early 1970s that Greenstein and Petersen (1973) reported
the detection of narrow cores within the hydrogen lines of DA white dwarfs.
Modeling these narrow cores, via NLTE effects convolved with rotational
broadening, allowed the possibility of measuring $v \sin i$.   FIrst results
using these cores, from Greenstein and Petersen (1973), Greenstein et al.
(1977) and Pilachowski \& Milkey (1984, 1987) all showed rotation velocities
at and below about 50 km/s -- which was seen largely as slow rotation under
the prevailing assumption of conservation of $j$.  More recent
high--resolution spectroscopy of white dwarfs by Heber et al. (1997) and
Koester et al. (1998) produced more accurate measurements than the earlier
work, but have not altered the conclusions drawn.  Heber et al. (1997) report
upper limits to rotation velocities to single non--magnetic white dwarfs. and
Koester et al. (1998) report rotation velocities for only a handful of
non--magnetic DA white dwarfs, including three ZZ Ceti stars.

Table 1 includes a representative sample of the upper limits and measurements
of $v \sin i$ for white dwarfs.  Note that except for the DAV (ZZ Ceti) stars,
only GD140 shows a rotation velocity that is not consistent with zero, and
that the uncertainties in the measurements are all of order 10 km/s.  This is
difficult spectroscopy - high resolution spectra are needed, and accurate and
reliable NLTE model atmospheres are essential for the analysis.

\begin{table} 
\caption{Representative white dwarf rotation rates.}

\begin{tabular}{crrccccr} \\ 
\tableline 
Star  & $M/M_{\odot}$ & $T_{\rm eff}$ & $\log g$ & $v$ [km/s] & $P_{\rm rot}$
[d]  & type & Ref \\ 
\tableline

\multicolumn{8}{c}{Spectroscopic determinations} \\

40 Eri B &   ~0.5  &16500 & 7.86  &    $\la$ 8  &$\ga$0.09  &  DA  &  1 \\
LP 207-7 &   ~0.05 &13600 & 7.76  &    $\la$18  &$\ga$0.04  &  "   &  1 \\ 
G 148-7  &    0.61 &15500 & 7.97  &    $\la$12  &$\ga$0.061 &  "   &  1 \\ 
HZ43     &    0.53 &49000 & 7.70  &    $\la$29  &$\ga$0.025 &  "   &  1 \\ 
G 1423-B2B &  0.52 &14000 & 7.83  &    $\la$9   &$\ga$0.081 &      &  1 \\ 
LB 253      &  -   &19200 & 7.8   &   20$\pm$20 & 0.03   &  "   & 3 \\ 
WD 1337+70  &  -   &21000 & 8.0   &   30$\pm$20 & 0.02   &  "   & 3 \\ 
GD 140      &  -   &23000 & 8.0   &   60$\pm$10 & 0.012  &  "   & 3 \\ 
W1346       &  -   &21500 & 8.0   &   10$\pm$10 & 0.06   &  "   & 4 \\ 
Grw+73 8031 &  -   &15400 & 8.0   &   40$\pm$20 & 0.015  &  "   & 4 \\ 
G29-38      & 0.6  &11600 & 8.1   &   45$\pm$5  & 0.014  & DAV  &  2 \\ 
GD165       & 0.6  &11950 & 7.9   &   27$\pm$7  & 0.024  & DAV  &  2 \\ 
L19-2       & 0.6  &12200 & 8.0   &   38$\pm$3  & 0.017  & DAV  &  2 \\ 

\tableline 

\multicolumn{8}{c}{Asteroseismic determinations}\\ 

G29-38     &  0.6  &11600 & 8.1   &     0.55  & 1.35  &  DAV  &  5 \\ 
GD165      &  0.6  &11950 & 7.9   &     0.18  & 4.20  &   "   &  6 \\ 
L19-2      &  0.6  &12200 & 8.0   &     0.63  & 1.10  &   "   &  7 \\ 
HS0507     &  0.6  &12000 & 8.0   &     0.47  & 1.54  &   "   &  8 \\ 
GD 358     &  0.61 &24000 & 8.0   &     0.606 & 1.20  &  DBV  &  9 \\ 
PG 0122+200 & 0.68 &75500 & 7.5   &     0.455 & 1.60  &  DOV  &  10 \\ 
PG 2131+066 & 0.62 &80000 & 7.5   &     3.487 & 0.21  &   "   &  11 \\ 
PG 1159-035 & 0.59 &140000 & 7.5   &     0.527 & 1.38  &   "   &  12 \\ 
RXJ 2117    & 0.56 &170000 & 6.0   &     0.627 & 1.16  &  PNNV & 13 \\ 
NGC 1501    & 0.55 & 81000 & 6.5   &     0.622 & 1.17  &  " & 14 \\ 
\tableline 
\tableline
\\

\end{tabular} 
1. Heber et al. (1997); 2. Koester et al. (1998); 3. Pilachowski \& Milkey
(1984); 4. Pilachowski \& Milkey (1987); 5. Kleinman et al. (1998); 6.
Bergeron et al. (1993); 7. O'Donaghue \& Warner (1987); 8. Handler et al.
(2002); 9. Winget et al. (1994); 10. O'Brien et al. (1998); 11. Kawaler et al.
(1995); 12. Winget et al. (1991); 13. Vauclair et al. (2002); 14. Bond et al.
(1996)

\end{table}

\subsection{Asteroseismology}

With few rotation velocities through spectroscopy, we must rely on a
relatively new technique to provide data to constrain models of angular
momentum evolution in highly evolved stars.  This new technique exploits the
observed nonradial $g-$mode pulsations of white dwarfs in several distinct
bands in the H--R diagram.  We see multimode nonradial pulsations in DA white
dwarfs with effective temperatures at about 12,000K, in DB white dwarfs at
about 25,000K, and in helium--poor C/O rich hot pre white dwarfs between
170,000K and 75,000K.  Comprehensive review of the many results for pulsating
white dwarfs can be found in many volumes (two personal favorites are Kawaler
2000 and Kawaler 1996, but see also Fontaine 2002).   Other papers in this
volume (i.e. Christiansen--Dalsgaard,  and Matthews) describe the effect of
slow rotation on the observed nonradial pulsation frequencies, so I only
briefly review a basic asymptotic result as it has been applied to pulsating
white dwarfs.

Nonradial modes of a given degree $\ell$ and radial overtone $n$ will have the
same pulsation frequency, independent of $m$, in a spherically symmetric star.
 Rotation can break this degeneracy.  For uniform rotation in white dwarfs
that are pulsating in $g-$modes,
\begin{displaymath}
\nu_{n \ell m} = \nu_{n \ell 0} + 
        \frac{m}{P_{\rm rot}} \left[ 1-\frac{1}{\ell (\ell +1)} \right],
\end{displaymath}
which is a limiting case where the pulsation motion in the horizontal
direction is much larger than the vertical direction.  So, if modes with all
values of $m$ are excited, the pulsation modes should be split into triplets
for $\ell=1$, into quintuplets for $\ell=2$, etc.  Also, the splitting of
these multiplets should be different for different values of $\ell$.  In one
star, PG~1159, Winget et al. (1991) show that these two effects are present: 
a sequence of triplet modes are split by equal amounts, and are nearly exactly
0.6 times the splitting seen in several quintuplets.  The theory works quite
well.

Rotational splitting has been seen in several white dwarfs (see Table 1). 
Asteroseismology clearly produces rotation frequencies of much higher
precision (especially at smaller rotation rates) than is possible through
spectroscopy.  Oddly, those three DAV stars that have rotation determinations
via spectroscopy show large discrepancies between the spectroscopic and
asteroseismic values.  Given the uncertainties in the spectroscopic values as
mentioned above, and the multiple determination of rotational splitting within
individual stars, the asteroseimsic values are probably more representative. 
We speculate that the broadening of the NLTE spectral line cores measured by
Koester et al. (1998) may be due to some other photospheric motions related to
the pulsation.

Is it possible that the interior of white dwarfs, where the pulsation
frequencies are determined, rotates much slower than the surface?  Kawaler,
Sekii, \& Gough (1999), explored the effects of differential rotation within
DOV and DBV white dwarfs.  They concluded that the surface layers dominate the
computed value of the rotational splitting in these stars.  The cooler DAV
stars are even more heavily influenced by the surface layers. Therefore the
seismological rotation rates should be similar to the surface rotational
broadening component of spectral lines.  (Though Winget et al. (1994)
suggested that the DBV GD358 showed evidence of differential rotation, Kawaler
et al. (1999) showed that the variation in rotational splitting with $n$ could
not be explained by a simple internal angular velocity gradient.) 

The conclusion that we draw about measured rotation rates for white dwarfs,
principally through asteroseismology, is that they rotate with periods
of about once per day, with a range from a few hours to a few days.  This is
strikingly similar to the average rotation rate measured in magnetic white
dwarfs, as reported by Wickramasinghe \& Ferrario (2000).  So, white dwarfs
rotate slowly (or is that quickly?).

\section{Evolutionary Models Using a Limiting Case for $j$
Redistribution}

\begin{figure}
\plotone{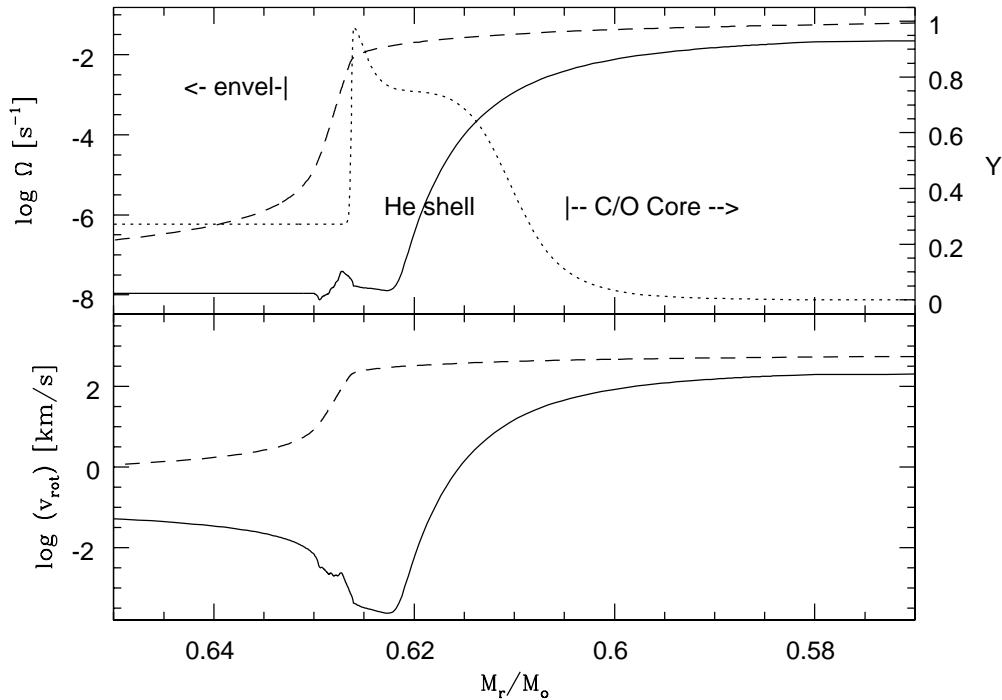}
\caption{The core boundary within an ADB model, following a thermal pulse. 
The abscissa is $M_{r}$ with the center to the right. Top panel:  angular
velocity $\Omega$ (solid line); the dashed line shows $\Omega$ if one assumes
constant $j$ everywhere throughout the evolution. The helium mass fraction
(dotted line) has been influenced by recent convective shell burning
during the thermal pulse.    Bottom panel: $v_{\rm rot}$ in the two limits.}
\end{figure}

With a bit of effort, evolutionary models that take us from the ZAMS through
the AGB phase can make more specific predictions about what the rotation rate
for white dwarfs might be.  Through helioseismology, we now know that the
interior of the Sun rotates as a solid body in the outer layers.  This (along
with ``common wisdom'') suggests that we should force convective regions
within stellar models to rotate as solid bodies, with the total angular
momentum within the convection zone as a conserved quantity.  As the
convection zone grows to incorporate more of the star, the angular momentum
``dredged up'' is distributed through the convection zone.  For a growing
surface convection zone, which is the dominant effect for AGB models, this
results in a draining of angular momentum from inner regions to outer layers. 
A retreating convection zone releases the newly radiative regions to rotate
with conserved local angular momentum.

To do this, we started with evolutionary sequences, begun on the ZAMS and
evolved through several thermal pulses on the AGB (Kawaler et al. 2002). We
then extracted the cores of these AGB models and evolved them into the white
dwarf domain. Throughout the calculation, we kept track of angular momentum
within the models using the prescription described above: conservation of $j$
in radiative regions, but solid body rotation within convective regions.

A sample ``seed model'' on the AGB is shown in Figure 1.  This model is
recovering from a recent thermal pulse.  Convective helium burning has
produced a marked plateau in the helium abundance profile at about
0.620$M_{\odot}$ from the center.  The transition zone from hydrogen to helium
at 0.626$M_{\odot}$ is the location of the recovering hydrogen burning shell,
which lies just below the extensive convective envelope.  The angular velocity
profile through this region reflects the flat rotation curve in the convection
zone.  The small rotational velocity inversion at the hydrogen shell caused by
its contraction that followed its release by the convective envelope,
and the rise in angular velocity within and below the helium burning shell and
into the compact C/O core.

If a star such as this leaves the AGB at this stage, it could produce a
hydrogen--rich planetary nebula (PN) central star should mass loss leave a
layer of approximately $10^{-4}M_{\odot}$ of hydrogen at the top of the core. 
The core would then contract to white dwarf dimensions, with the outer layers
spinning up.  The resulting DA white dwarf would then have a rotation profile
resembling that shown in Figure 2, which is the 0.626$M_{\odot}$ white dwarf
remnant from the model shown in Figure 1.  Figure 2 plots quantities versus
the log of the surface mass fraction to expand features around the composition
transition zones.

\begin{figure}
\plotone{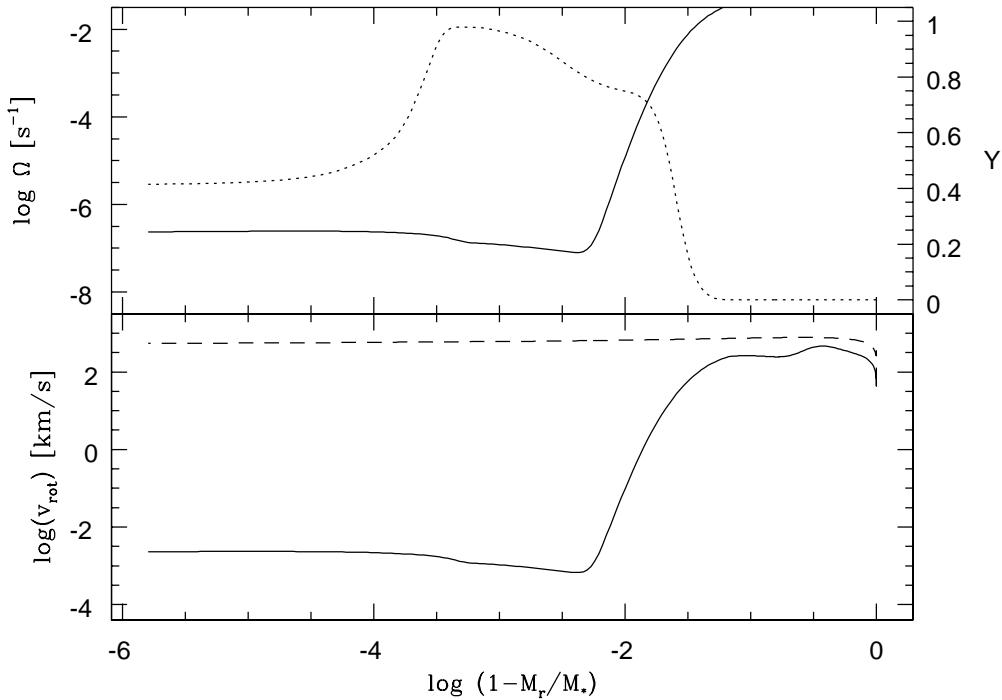}
\caption{Rotation profile within a hot hydrogen--rich DA white dwarf
model. Properties are plotted with respect to the log of the surface mass
fraction, with the surface to the left and the center to the right. Lines are
as described for Figure 1.}
\end{figure}
The situation for DB white dwarfs or PG~1159 stars is less desperate.  Figures
1 and 2, suggest that loss of mass into the plateau of the helium profile
($10^{-2}$ $M_{*}$ below the surface) will expose material with a much larger
rotation velocity.  Figure 3 shows a hot PG~1159 white dwarf model that
represents the remnant of the model in Figure 1 should mass loss reach to the
base of the plateau region.   The rotation rate in the surface layers is quite
close to the observed values. 

These conclusions do depend on the initial angular momentum (and initial main
sequence mass).  We did not include any form of diffusive angular momentum
transport, which could allow angular momentum to leak out of the core and spin
up the surface layers.  This would happen on time scales that are much longer
than the time scales for evolution of hot white dwarfs.  Typical angular
momentum diffusion time scales are of order $10^{7}$ years, which is 100 times
longer than the time between thermal pulses; therefore, the angular momentum
distribution within AGB stars is dominated by the convective envelope motion
and convection within the thermally pulsing shells. Once a white dwarf is
formed, takes much less than $10^6$ years to cross the H--R diagram and evolve
down the top of the white dwarf cooling track.  

Note that the observed rotation rates in pulsating white
dwarfs, $\log v = -0.6$ or $\log \Omega = -4.3$, are much faster than the
model.  Spectroscopic upper limits of 10 km/s are orders of magnitude larger
than this model predicts.  

\begin{figure}
\plotone{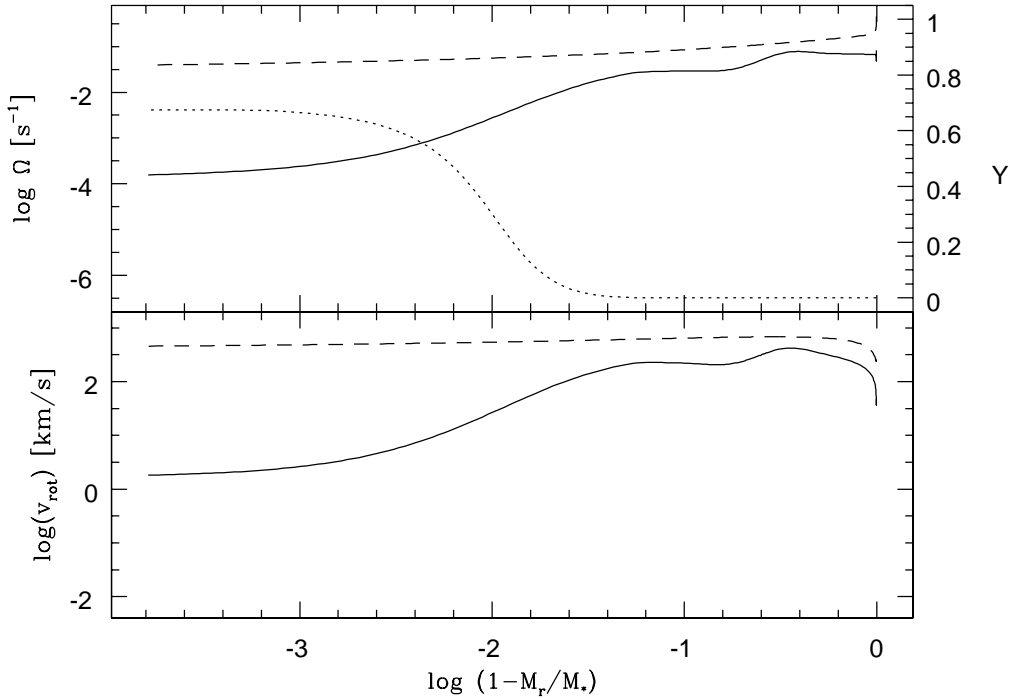}
\caption{Same as Figure 2 but for a DB white dwarf model.}
\end{figure}
Stars reach the DAV instability strip about $10^{8}$ years past the AGB; by
then angular momentum diffusion may act to spin up the outer layers of DA
white dwarfs.  Langer et al. (1999) computed models up the AGB that included
rotation and angular momentum redistribution.  Though concentrating on
rotationally induced mixing and its influence on nucleosynthesis, they
extrapolate their results into the white dwarf phase to suggest that white
dwarf rotation velocities of about 25 km/s might be expected.   This
conclusion does depend on the precise mass of the remnant (as indicated in
our simpler models above), and is is clearly faster than seen in the pulsating
DA white dwarfs.

\section{Conclusions, and Speculations}

Spectroscopic measurements of $v \sin i$ in single white dwarfs provide
weak constraints on their rotation.  Asteroseismology provides an accurate
measurement, though only for those that are observed to pulsate.  As the
pulsators are representative of the class of stars, seismic rotation
velocities can give some insight into the angular momentum evolution during
the late stages of stellar evolution.

Reasonable assumptions about angular momentum transport within evolving
stellar interiors points to several preliminary conclusions. Clearly, $j$ is
drained from the outer core into the convective envelope on the AGB.  As a
result, DA white dwarfs are left with material with relatively low $j$, as
their layers are repeatedly coupled to the stellar surface by convection. Thus
they should (initially) rotate with periods much longer than one day.  DB (and
DO) white dwarfs have surfaces that were deeper than the deepest extent of the
AGB envelope, and therefore retain higher $j$ -- and should be rotating
(initially) significantly faster than DA white dwarfs.  Future work will
concentrate on models of white dwarfs that include these angular velocity
profiles, testing to see if the differential rotation can produce the observed
changes in rotational splitting with overtone $n$ -- in the manner described
by Kawaler et al. (1999).

Finally, we point out that the rapid evolution along the AGB drives steep
angular velocity gradients at the core/envelope boundary.  Such steep
gradients can perhaps drive a magnetic dynamo, and produce a shaping mechanism
for planetary nebulae.  This possibility has been investigated by several
groups (see, for example, Blackman et al. 2001).

\acknowledgments

This work was supported, in part, by NASA ATP Grant NAG5-8352 to Iowa State
University.  Shelbi Hostler was instrumental in carrying out many of the
calculations reported here, and Hugh Van Horn provided the stimulus and
encouragement to look into this problem.


\begin{references}

\reference 
Bergeron, P., et al. (the WET collaboration) 1993, \aj\/  106, 1987
\reference
Blackman, E., Frank, A., Markiel, J.A., Thomas, J.H., \& Van Horn, H.M. 2001,
\nat\/ 409, 485
\reference
Bond, H., et al. 1996, \aj\/  112, 2699 
\reference
Fontaine, G. 2002, in: M. Cunha (ed.), Asteroseismology Across the H--R
Diagram, (San Francisco, ASP), in press 
\reference
Greenstein J.L., Boksenberg A., Carswell R., \& Shortridge K. 1977, \apj\/
212, 186
\reference 
Greenstein, J.L. \& Petersen, D.M. 1973, \aap \/ 29, 23 
\reference 
Handler, G., Romero-Colmenero, E., \& Montgomery, M. 2002, \mnras\/  335, 399
\reference 
Heber, U., Napiwotski, R., \& Reid, I.N. 1997, \aap \/ 323, 819
\reference
Kawaler, S., Hostler, S., \& Burkett, J. 2002, in: R. Silvotti \& D. DiMartini
(eds.), NATO Advance Research Workshop: The 13th European Workshop on White
Dwarfs, (Dordrecht: Kluwer), in press
\reference
Kawaler, S. 2000, in: C. Ibanoglu (ed.), NATO Advanced Study Institute:
Variable Stars as Astrophysical Tools, (Dordrecht: Kluwer), 511
\reference
Kawaler, S., Sekii, T., \& Gough, D. 1999, \apj\/ 516, 349
\reference
Kawaler, S. 1996, in: G. Meynet \& D. Schaerer (eds.), Stellar Remnants: Saas
Fee Advanced Course 25, (Berlin: Springer), 1
\reference 
Kawaler, S. et al. (the WET collaboration) 1995, \apj\/  450, 350
\reference 
Kawaler, S. 1987, \pasp\/ 99, 1322 
\reference 
Kleinman, S. et al. (the WET collaboration) 1998, \apj\/  495, 424 
\reference 
Koester, D., Dreizler, S., Weidemann, V., \& Allard, N. 1998, \aap\/  338, 612 
\reference
Langer, N., Heger, A., Wellstein, S., \& Herwig, F. 1999, \aap\/ 346, L37
\reference
O'Brien, M.S. et al. (the WET collaboration) 1998, \apj\/  495, 458 
\reference
O'Donoghue, D. \& Warner, B. 1987, \mnras\/  228, 949 
\reference 
Pilachowski, C. \& Milkey, R. 1984, \pasp\/  96, 821 
\reference 
Pilachowski, C. \& Milkey, R. 1987, \pasp\/  99, 836 
\reference 
Roxburgh, I.W. 1965, \zap\/  62, 134
\reference 
Spruit, H. 1998, \aap\/  333, 603 
\reference 
Vauclair, G. et al. (the WET collaboration) 2002, \aap\/ 381, 122
\reference
Wickramasinghe, D. \& Ferrario, L. 2000, \pasp\/ 112, 873
\reference 
Winget, D. et al. (the WET collaboration) 1991, \apj\/  378, 326 
\reference 
Winget, D. et al. (the WET collaboration) 1994, \apj\/  430, 839

\end{references}
\end{document}